\documentstyle[11pt]{article}
\begin{document}

\begin{flushright}
gr-qc/9504004\\
UMDGR-95-114\\
\end{flushright}

\begin{center}
{\LARGE\bf Thermodynamics of Spacetime:\\
\vskip 2mm
 The Einstein Equation of State}
\vskip .5cm
Ted Jacobson
\vskip .5cm
{{\it Department of Physics, University of Maryland\\
                          College Park, MD 20742-4111, USA}\\
             {\tt jacobson@umdhep.umd.edu}}

\end{center}
\vskip .5cm

\begin{abstract}
The Einstein equation is derived from the
proportionality of entropy and horizon area
together with the fundamental relation
$\delta Q=TdS$ connecting heat, entropy, and temperature.
The key idea is to demand that this relation hold for all the
local Rindler causal horizons through each spacetime point, with
$\delta Q$ and $T$ interpreted as the energy flux and Unruh temperature
seen by an accelerated observer just inside the horizon.
This requires that gravitational lensing by matter energy
distorts the causal structure of spacetime in just such a
way that the Einstein equation holds.
Viewed in this way, the Einstein equation is an equation of state.
This perspective suggests that it may be
no more appropriate to canonically quantize the Einstein equation than
it would be to quantize the wave equation for sound in air.

\end{abstract}

\vskip 1cm

The four laws of black hole mechanics, which are analogous to those
of thermodynamics, were originally
derived from the classical Einstein equation\cite{barcarhaw}.
With the discovery of the quantum Hawking radiation\cite{hrad},
it became clear that the analogy is in fact an identity.
How did classical General Relativity know that horizon area
would turn out to be a form of entropy, and that surface gravity
is a temperature? In this letter I
will answer that question by turning the logic around
and deriving the Einstein equation
from the proportionality of entropy and horizon
area together with the
fundamental relation $\delta Q=TdS$
connecting heat $Q$, entropy $S$, and temperature $T$.
Viewed in this way,
the Einstein equation is an equation of state. It is
born in the thermodynamic limit as a relation between thermodynamic
variables, and its validity is seen to depend on the existence of
local equilibrium conditions.
This perspective suggests that it may be no more
appropriate to quantize the Einstein equation than
it would be to quantize the wave equation for sound
in air.

The basic idea can be illustrated by thermodynamics
of a simple homogeneous system. If one knows the entropy
$S(E,V)$ as a function of energy and volume,
one can deduce the equation of state from
$\delta Q=T dS$. The first law of thermodynamics
yields $\delta Q=dE+pdV$, while differentiation yields
the identity
$dS=(\partial S/\partial E)\, dE + (\partial S/\partial V)\, dV$.
One thus infers that $T^{-1}=\partial S/\partial E$ and
that $p=T(\partial S/\partial V)$. The latter equation is the
equation of state, and yields useful information if the
function $S$ is known. For example, for weakly interacting
molecules at low density, a simple counting argument yields
$S=\ln(\# {\rm ~accessible~states})\propto \ln V + f(E)$ for some
function $f(E)$. In this case, $\partial S/\partial V\propto V^{-1}$,
so one infers that $pV\propto T$, which is the equation of state
of an ideal gas.

In thermodynamics, heat is energy that flows between degrees
of freedom that are not macroscopically observable.
In spacetime dynamics, we shall define
heat as energy that flows across a causal horizon. It can be felt
via the gravitational field it generates, but its particular form
or nature is unobservable from outside the horizon.
For the purposes of this definition
it is not necessary that the horizon be a black hole event
horizon. It can be simply the boundary of the past of any
set ${\cal O}$  (for ``observer").
This sort of horizon is a null hypersurface
(not necessarily smooth) and, assuming cosmic censorship, it is
composed of generators which are
null geodesic segments emanating backwards in time
from the set ${\cal O}$.
We can consider a kind of local gravitational
thermodynamics associated with such causal
horizons, where the ``system"
is the degrees of freedom beyond the horizon.
The outside world is separated from the system not by a
diathermic wall, but by a causality barrier.

That causal horizons should be associated with entropy is suggested
by the observation that they hide information\cite{bek}.
In fact, the overwhelming majority of the
information that is hidden resides
in correlations between vacuum fluctuations just inside and outside
of the horizon\cite{tangle}.
Because of the infinite number of short wavelength field degrees of
freedom near the horizon, the associated ``entanglement entropy" is divergent
in continuum quantum field theory. If, on the other hand, there is a
fundamental cutoff length $l_c$, then the entanglement entropy
is finite and proportional to the horizon area in units of $l_c^2$,
as long as the radius
of curvature of spacetime is much longer than $l_c$.
(Subleading dependence on curvature
and other fields induces subleading terms in the gravitational
field equation.)
We shall thus assume for most of this letter
that the entropy is proportional to horizon area.
Note that the area is an extensive quantity for a horizon,
as one expects for entropy\cite{holo}.

As we will see, consistency with thermodynamics requires that
$l_c$ must be of order the Planck length ($10^{-33}$ cm). Even
at the horizon of a stellar mass black hole, the radius of curvature
is $10^{38}$ times this cutoff scale. Only near the big bang or
a black hole singularity, or in the final stages of evaporation
of a primordial black hole, would such a vast separation of scales
fail to exist. Our analysis relies heavily on this circumstance.

So far we have argued that energy flux across a causal horizon
is a kind of heat flow, and that entropy of the system beyond
is proportional to the area of that horizon.
It remains to identify the temperature of the system
into which the heat is flowing. Recall that the origin of
the large entropy is the vacuum fluctuations of quantum fields. According
to the Unruh effect\cite{unruh},
those same vacuum fluctuations have a thermal
character when seen from the perspective of a uniformly accelerated
observer. We shall thus take the temperature of the system to be the
Unruh temperature associated with such an observer
hovering just inside the horizon. For consistency, the same
observer should be used to measure the energy flux that defines the
heat flow. Different accelerated observers will obtain
different results. In the limit that the accelerated worldline approaches
the horizon the acceleration diverges, so the Unruh temperature
and energy flux diverge, however their ratio approaches a finite limit.
It is in this limit that we analyse the thermodynamics, in order to
make the arguments as local as possible.

Up to this point we have been thinking of the system as
defined by any causal horizon. However, in general, such
a system is not in ``equilibrium" because the horizon
is expanding, contracting, or shearing. Since we wish to
apply equilibrium thermodynamics, the system is further
specified as follows. The equivalence principle is invoked to
view a small neighborhood of each spacetime point $p$
as a piece of flat spacetime. Through $p$ we consider a small
spacelike 2-surface element ${\cal P}$ whose past directed
null normal congruence to one side (which we call the ``inside")
has vanishing expansion and shear at $p$.  It is always
possible to choose ${\cal P}$ through $p$ so that the expansion and
shear vanish in a first order neighborhood of $p$. We call
the past horizon of such a $P$ the ``local Rindler horizon of
${\cal P}$", and we think of it as defining a system---the part of
spacetime beyond the Rindler horizon---that is instantaneously
stationary (in ``local equilibrium") at $p$.
Through any spacetime point there are local Rindler horizons in
all null directions.

The fundamental principle at play in our analysis is this:
The equilibrium thermodynamic relation $\delta Q=T dS$,
as interpreted here in terms of energy flux and area of
local Rindler horizons, can only be satisfied if gravitational
lensing by matter energy distorts the causal structure of
spacetime in just such a way that the Einstein equation holds.
We turn now to a demonstration of this claim.

First, to sharpen the the above definitions of temperature and heat,
note that in a small neighborhood of any spacelike 2-surface
element ${\cal P}$ one has an approximately flat
region of spacetime with the usual Poincar\'e symmetries.
In particular, there is an approximate Killing field $\chi^a$
generating boosts orthogonal to ${\cal P}$ and vanishing at ${\cal P}$.
According to the Unruh effect\cite{unruh}, the Minkowski
vacuum state of quantum fields---or any state at very short distances---
is a thermal state with respect to the boost hamiltonian
at temperature $T=\hbar \kappa/2\pi$, where
$\kappa$ is the acceleration of the Killing orbit on which
the norm of $\chi^a$ is unity (and we employ units with the
speed of light equal to unity.) The heat flow is to be defined by the
boost-energy current of the matter, $T_{ab}\chi^a$, where
$T_{ab}$ is the matter energy-momentum tensor.
Since the temperature and heat flow scale the same way under
a constant rescaling of $\chi^a$, this scale ambiguity will
not prevent us from inferring the equation of state.

Consider now any local Rindler horizon through a spacetime point $p$.
(See Fig. 1.)
Let $\chi^a$ be an approximate local boost Killing field
generating this horizon, with the direction of $\chi^a$ chosen to
be future pointing to the ``inside" past of ${\cal P}$. We assume
that all the heat flow across the horizon is (boost) energy
carried by matter. This heat flux to the past of ${\cal P}$ is given by
\begin{equation}
\delta Q=\int_{\cal H} T_{ab}\chi^a d\Sigma^b.
\end{equation}
(In keeping with the thermodynamic limit, we assume
the quantum fluctuations in $T_{ab}$ are negligible.)
The integral is over a pencil of generators of the ``inside"
past horizon ${\cal H}$ of ${\cal P}$.
If $k^a$ is the tangent vector to the horizon generators for
an affine parameter $\lambda$ that vanishes at ${\cal P}$
and is negative to the past of ${\cal P}$, then
$\chi^a=-\kappa\lambda k^a$ and $d\Sigma^a=k^a d\lambda d{\cal A} $,
where $d{\cal A}$ is the area element on a cross section of the
horizon. Thus the heat flux can also be written as
\begin{equation}
\delta Q=-\kappa\int_{\cal H} \lambda \, T_{ab}k^a k^b \,
d\lambda d{\cal A}.
\label{dQ}
\end{equation}

Assume now that the entropy is proportional to the horizon
area, so the entropy variation associated with a piece of the
horizon satisfies $dS=\eta\, \delta{\cal A}$, where $\delta{\cal A}$ is
the area variation of a cross section of a pencil of generators
of ${\cal H}$. The dimensional constant $\eta$ is
undetermined by anything we have said so far
(although given a microscopic theory of spacetime structure
one may someday be able to compute $\eta$ in terms of a
fundamental length scale.)
The area variation is given by
\begin{equation}
\delta{\cal A}=\int_{\cal H} \theta \, d\lambda d{\cal A},
\end{equation}
where $\theta$ is the expansion of the horizon generators.

The content of $\delta Q=T dS$ is essentially to require that
the presence of the energy flux is associated with a focussing
of the horizon generators. At ${\cal P}$ the local Rindler horizon
has vanishing expansion, so the focussing to the past of
${\cal P}$ must bring an expansion to zero at just the right rate so that
the area increase of a portion of the horizon will be
proportional to the energy flux across it.
This requirement imposes a condition on the curvature of
spacetime as follows.

The equation of geodesic deviation applied to the null geodesic
congruence generating the horizon yields the Raychaudhuri equation
\begin{equation}
{d\theta\over d\lambda}=-{1\over2}\theta^2-\sigma^2-R_{ab}k^ak^b,
\end{equation}
where $\sigma^2=\sigma^{ab}\sigma_{ab}$ is the square of the
shear and $R_{ab}$ is the Ricci tensor.
We have chosen the local Rindler horizon to be instantaneously
stationary at ${\cal P}$,
so that $\theta$ and $\sigma$ vanish at ${\cal P}$. Therefore
the $\theta^2$ and $\sigma^2$ terms are higher order contributions
that can be neglected compared with the last term when integrating to
find $\theta$ near ${\cal P}$. This integration yields
$\theta=-\lambda\, R_{ab} k^a k^b $ for sufficiently
small $\lambda$. Substituting this into the equation
for $\delta {\cal A}$ we find
\begin{equation}
\delta A=-\int_{\cal H} \lambda \, R_{ab}k^a k^b \, d\lambda d{\cal A} .
\label{dA}
\end{equation}

With the help of (\ref{dQ})  and (\ref{dA}) we can now see that
$\delta Q=TdS=(\hbar\kappa/2\pi)\eta\, \delta A $ can only be valid if
$T_{ab}k^ak^b=(\hbar\eta/2\pi)\; R_{ab} k^ak^b$
for all null $k^a$, which implies that
$(2\pi/\hbar\eta)T_{ab}= R_{ab} + f g_{ab}$
for some function $f$.
Local conservation of energy and momentum implies that
$T_{ab}$ is divergence free and therefore, using the contracted
Bianchi identity, that
 $f=-R/2 +\Lambda$ for some
constant $\Lambda$. We thus deduce that the Einstein equation
holds:
\begin{equation}R_{ab}-{1\over 2}R g_{ab}+\Lambda g_{ab}=
{2\pi\over\hbar\eta}T_{ab}.
\end{equation}
The constant of proportionality $\eta$
between the entropy and the area
determines Newton's constant as $G=(4\hbar\eta)^{-1}$,
which identifies the length $\eta^{-1/2}$ as twice
the Planck length $(\hbar G)^{1/2}$. The undetermined
cosmological constant
$\Lambda$ remains as enigmatic as ever.

Changing the assumed entropy functional would change the implied
gravitational field equations. For instance, if the entropy density
is given by a polynomial in the Ricci scalar
$\alpha_0+\alpha_1 R+...$, then $\delta Q=TdS$ will imply
field equations arising from a Lagrangian polynomial in the
Ricci scalar\cite{poly}. It is an interesting question
what ``integrability" conditions must an entropy density satisfy
in order for $\delta Q=TdS$ to hold for all local Rindler horizons.
It seems likely that the requirement is that the entropy density
arises from the variation of a generally covariant action just as it
does for black hole entropy. Then the implied field equations will
be those arising from that same action.

Our thermodynamic derivation of the Einstein equation of state
presumed the existence of local equilibrium conditions
in that the relation $\delta Q=T dS$ only applies
to variations between nearby states of local thermodynamic
equilibrium. For instance, in free expansion of a gas, entropy
increase is not associated with any heat flow, and this relation
is not valid. Moreover, local temperature and entropy are not
 even well defined away from equilibrium.
In the case of gravity, we chose our systems to be defined by
local Rindler horizons, which are instantaneously stationary, in
order to have systems in local equilibrium.
At a deeper level, we also assumed the usual form of short
distance vacuum fluctuations in quantum fields when we motivated
the proportionality of entropy and horizon area and the use of
the Unruh acceleration temperature.
Viewing the usual vacuum as a zero temperature thermal
state\cite{sciama}, this also amounts to a sort of local
equilibrium assumption. This deeper
assumption is probably valid only in some extremely good approximation.
We speculate that out of equilibrium vacuum fluctuations
would entail an ill-defined spacetime metric.

Given local equilibrium conditions, we have in the Einstein equation
a system of local partial differential equations that is time reversal
invariant and whose solutions include propagating waves. One might think of
these as analogous to sound in a gas propagating as an adiabatic
compression wave. Such a wave is a travelling disturbance
of local density, which propagates via a myriad of incoherent collisions.
Since the sound field is only a statistically defined observable
on the fundamental phase space of the multiparticle system,
it should not be canonically quantized as if it were a fundamental
field, even though there
is no question that the individual molecules are quantum mechanical.
By analogy, the viewpoint developed here suggests that it may
not be correct to canonically quantize the Einstein equations,
even if they describe a phenomenon that is ultimately quantum mechanical.

For sufficiently high sound frequency or intensity one knows that
the local equilibrium condition breaks down, entropy increases,
and sound no longer propagates in a time reversal invariant manner.
Similarly, one might expect that sufficiently high frequency or
large amplitude disturbances of the gravitational field would
no longer be described by the Einstein equation, not because
some quantum operator nature of the metric would become relevant,
but because the local equilibrium condition would fail.
It is my hope that, by following this line of inquiry, we
shall eventually reach an understanding
of  the nature of ``non-equilibrium spacetime".

\vskip 1cm
I am grateful to S. Corley, J.C. Dell, R.C. Myers, J.Z. Simon, and
L. Smolin for helpful comments on the presentation in earlier
drafts of this letter.
This work was supported in part by NSF grant PHY94-13253.

\vskip 1cm
\noindent FIGURE 1: Spacetime diagram showing the heat flux $\delta Q$
across the local Rindler horizon $\cal H$ of a 2-surface element
$\cal P$. Each point in the diagram represents a two dimensional
spacelike surface. The hyperbola is a uniformly accelerated worldline,
and $\chi^a$ is the approximate boost Killing vector on $\cal H$.


\begin{thebibliography}{999999}
\bibitem
{holo}Another
argument that might be advanced in support of the
proportionality of entropy and area comes from the holographic
hypothesis\cite{thooft,susskind},
ie, the idea that the state of the part of the
universe inside a spatial region can be fully specified
on the boundary of that region. However, currently the
primary support for this hypothesis comes from black hole
thermodynamics itself. Since we are trying to account for the
occurrence of thermodynamic-like laws for classical black
holes it would therefore be circular to invoke this argument.
\bibitem
{poly}If the entropy density on the horizon takes the form
$e^\rho d{\cal A}$,
with $\rho$ a function of the Ricci scalar,
one finds that $\delta Q=T dS$ implies
$(2\pi/\hbar\eta)T_{ab}= R_{ab} -\nabla_a\nabla_b\rho + f g_{ab}$
for some function $f$\cite{secondlaw}.
Imposing the local energy conservation equation
$\nabla^a T_{ab}=0$ then allows one to solve uniquely for $f$
up to a constant.
\end{thebibliography}
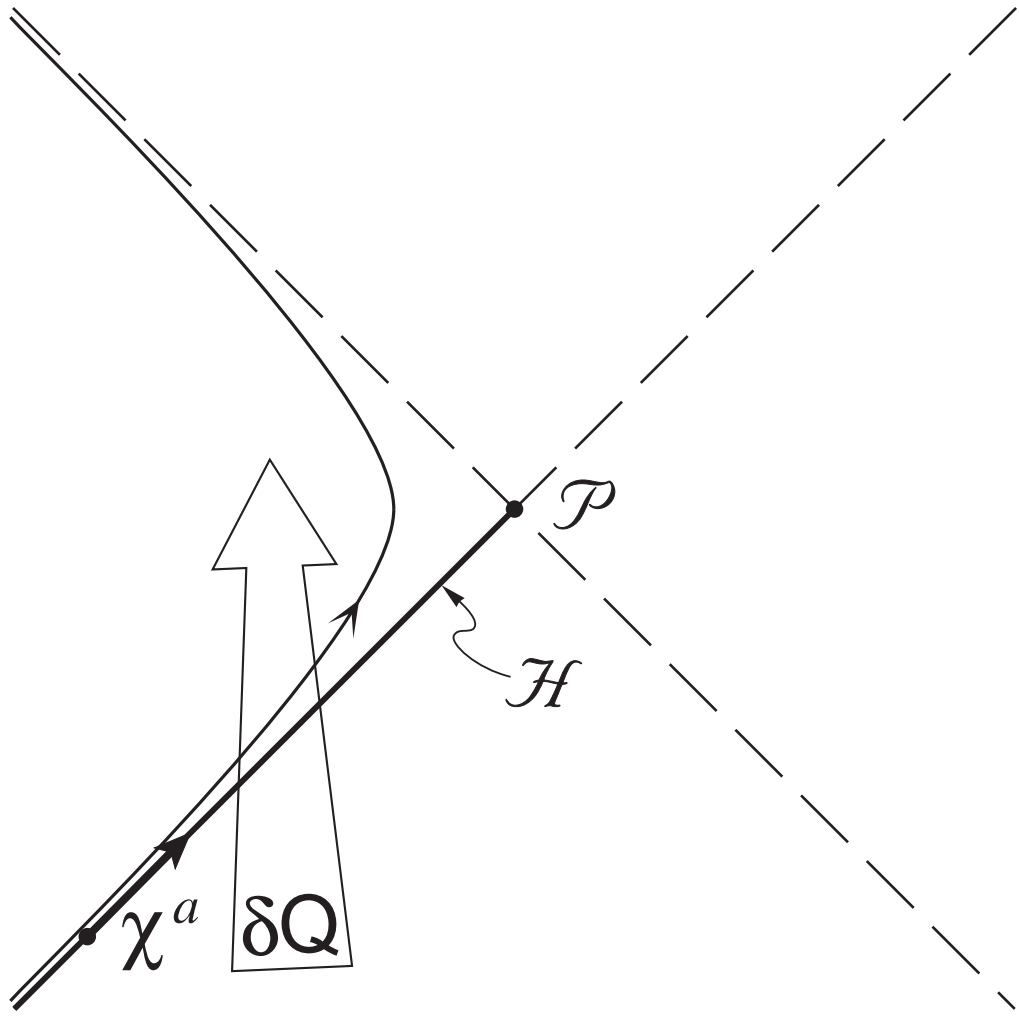
\end{document}